\documentclass{PoS}

\usepackage{lineno}
\usepackage{graphicx}
\usepackage[loose]{units}
\usepackage{upgreek}
\usepackage{color}
\usepackage[usenames,dvipsnames]{xcolor}
\usepackage{amsmath,amssymb,amsbsy}
			
\newcommand{\pT}{p_{\mbox{\tiny T}}}
\newcommand{\sNN}{$\sqrt{s_{\mbox{\tiny NN}}}$ }
\newcommand{\s}{$\sqrt{s}$ }

\newcommand{\Pb}{{\mbox{Pb--Pb}} }
\newcommand{\pPb}{{\mbox{p--Pb}} }

\title{Light neutral meson production in the era of precision physics at the LHC}

\ShortTitle{Light neutral meson production in the era of precision physics at the LHC}

\author{\speaker{Mike Sas}\thanks{for the ALICE Collaboration}\\
        University of Utrecht \& Nikhef, Netherlands\\
        E-mail: \email{mike.sas@cern.ch}}

\abstract{
The production of light neutral mesons in different collision systems is interesting for a variety of reasons: In nucleus-nucleus (AA) collisions the measurements provide important information on the energy loss of partons traversing the Quark-Gluon Plasma (QGP) which is formed in heavy-ion collisions at the LHC. In proton--proton (pp) collisions, neutral mesons allow us to test with high precision the predictions of perturbative QCD and other model calculations, and also serve as a reference for pA and AA collisions. In pA collisions, cold nuclear matter effects are studied.

In the ALICE experiment, which is dedicated to the study of the QGP, neutral mesons can be detected via their decay to two photons. The latter can be reconstructed using the two calorimeters EMCal and PHOS or via conversions in the detector material.

Combining state-of-the-art reconstruction techniques with the large data sample delivered by the LHC in Run 2 gives us the opportunity to enhance the precision of our measurements. In these proceedings, an overview of neutral meson production in pp, p--Pb and Pb--Pb collisions at LHC energies, as measured with the ALICE detector is presented.
}

\FullConference{
European Physical Society Conference on High Energy Physics - EPS-HEP2019 -\\
			10-17 July, 2019\\
			Ghent, Belgium}

\begin{document}
\section{Introduction}

In the field of heavy-ion physics we are faced with interesting questions such as: what are the different particle production mechanisms across different system sizes? can we find the onset of the QGP in heavy-ion collisions? and is there a QGP droplet formed in small collision systems?~\cite{Wilke:2018}
In proton--proton collisions the particle production mechanism should be dominated by the fragmentation of high momentum partons into jet-like structures. In collisions of heavy nuclei such as \Pb the production of particles is expected to be dominated by the hadronisation of the QGP. Studying the particle production mechanisms is thus key to understand the physics governing both small and large systems.

Identified hadron spectra are a good probe to study both the production mechanisms in pp collisions~\cite{Sassot:2010}, as well as the parton energy loss in heavy-ion collisions. Among these identified hadrons are the neutral pion ($\pi^{0}$) and $\eta$ meson, which are very abundant and have large branching ratios into two photons, making them suitable probes to study particle production. In addition, measuring neutral mesons grants the possibility of extracting the direct photon yield which come as an excess yield above the photons from hadronic decays, which probes the temperature of the QGP~\cite{Shen:2013vja}.

We present the invariant yield of neutral mesons in pp at \s$ = 5$ TeV, and \pPb and \Pb collisions at \sNN$ = 5.02$ TeV, with the ALICE detector. The measurements are compared to event generators and model calculations.

\section{Method}
\subsection{Photon reconstruction}
The neutral mesons are measured using the ALICE detector~\cite{Aamodt:2008zz} via the two photon decay channel. The photons are reconstructing using the photon conversion method (PCM), and with the calorimeters PHOS and EMCal.
With PCM, the photons that convert in the detector material to an electron and positron pair are reconstructed using the ITS and TPC detectors. With a conversion probability of 8\%, this method is limited in statistical precision but it profits greatly from the high momentum resolution of the ALICE central barrel.\\
The calorimeters are situated outside the inner detectors and are able to measure the photons by absorbing their full energy in their calorimeter towers.
The PHOS calorimeter consists of lead tungstate crystals with a cell size of 2.2 cm $\times$ 2.2 cm at a radius of 4.6 m from the collision region.
The EMCal is a Pb-scintillator sampling calorimeter with a cell size of 6 cm $\times$ 6 cm at a radius of 4.28 m from the collision region. The PHOS has a slightly better energy resolution but the EMCal has a larger acceptance. Both calorimeters have trigger capabilities to select events with high energetic clusters.\\
Combining these different photon reconstruction techniques allows us to reduce the statistical and systematic uncertainties of neutral meson reconstruction.

\subsection{Meson reconstruction}
The neutral mesons are reconstructed as follows.
First, the photons are reconstructed and the invariant mass of every photon pair is calculated. The neutral meson yield is situated on top of a combinatorial background. Second, the meson raw yield is obtained by integrating the invariant mass distributions around their corresponding mass, after subtracting the combinatorial and remaining background. Third, the raw yield is corrected for efficiency, acceptance, and feed-down from secondaries. At last, the different reconstruction methods are combined into a single measurement.

\begin{figure}
	\centering
	\includegraphics[width=0.49\textwidth]{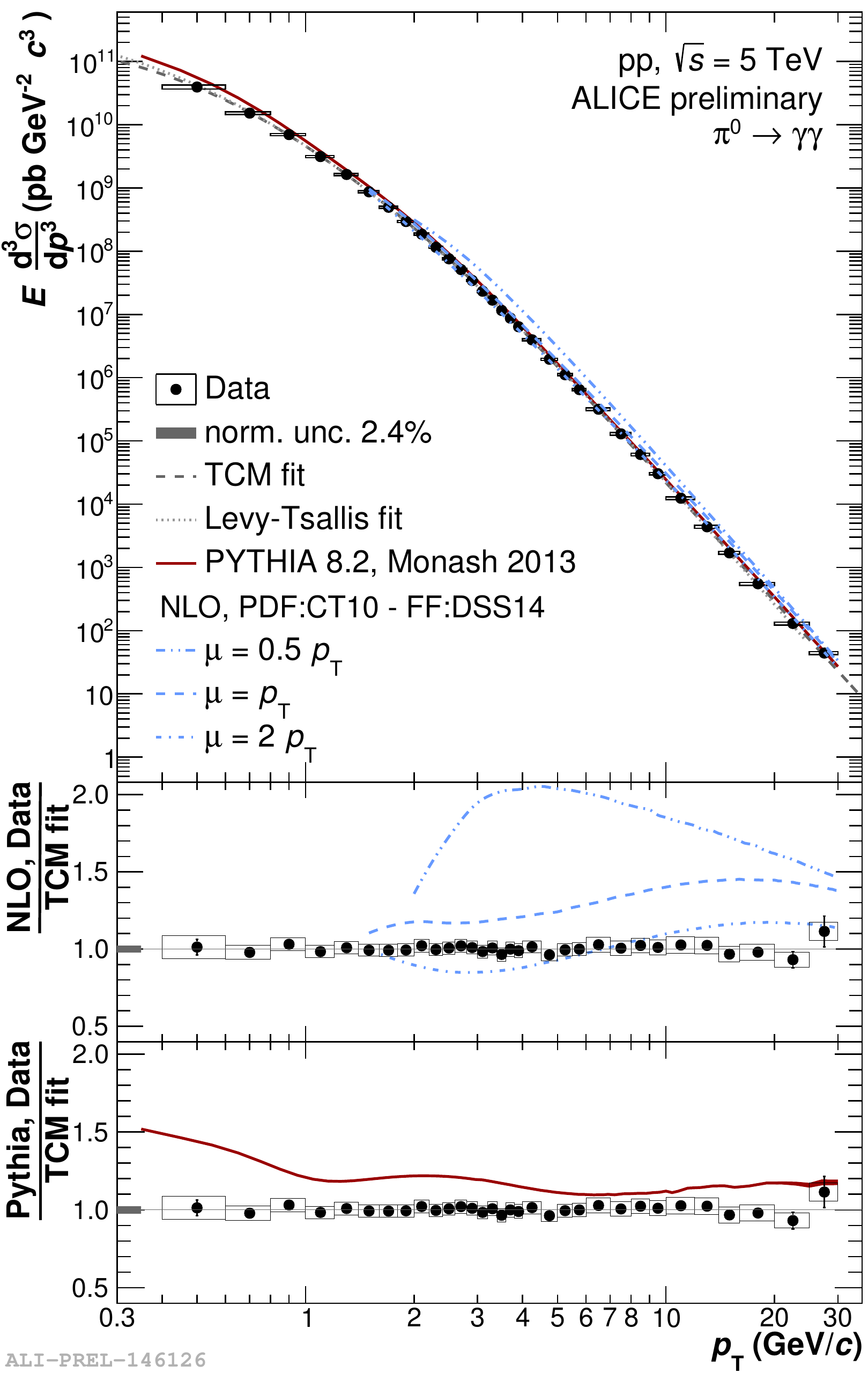}
	\includegraphics[width=0.49\textwidth]{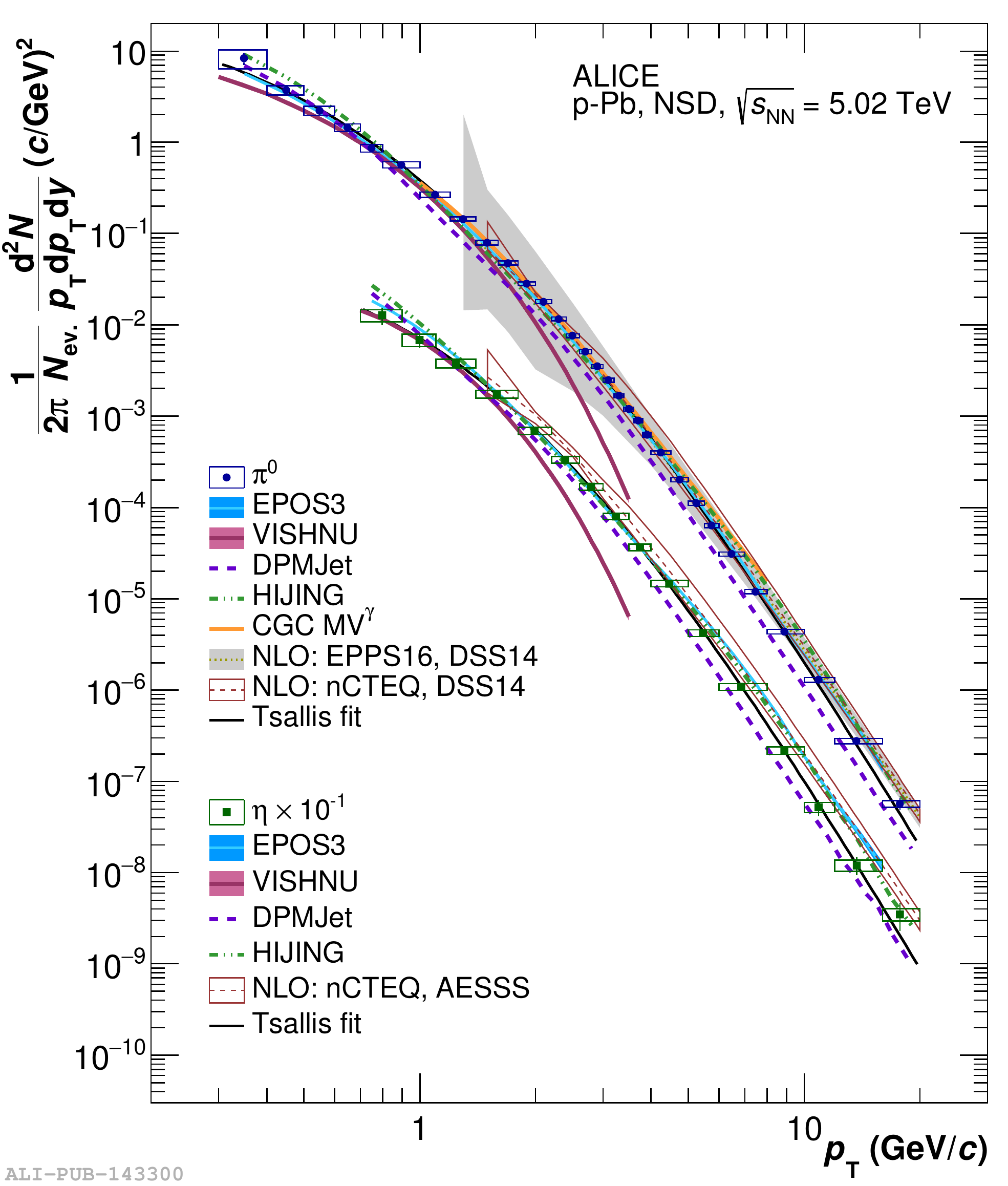}
	\caption{The production of neutral pions in pp collisions at \s$ = 5$ TeV (left), and neutral pion and eta mesons in \pPb collisions at \sNN$ = 5.02$ TeV (right)~\cite{AlicepPb:2018}.}
	\label{pp_pPb_production}
\end{figure}

\section{Results}
\subsection{pp collisions}

Figure \ref{pp_pPb_production} (left) shows the invariant yield of neutral pions as function of transverse momentum in pp collisions at \s$ = 5$ TeV. It is measured using the PCM and the calorimeters PHOS and EMCal.
It covers $0.4<\pT<30$ GeV/$c$, and is compared to the PYTHIA event generator~\cite{Pythia:2014} and NLO pQCD predictions, which both over-predict the $\pi^{0}$ production. This spectrum is used also as a reference for the yield of neutral pions in \pPb and \Pb collisions. In the future, more differential studies would be able to disentangle if this difference comes from either jet production or the underlying event.

\subsection{p--Pb collisions}

\begin{figure}
	\centering
	\includegraphics[width=0.49\textwidth]{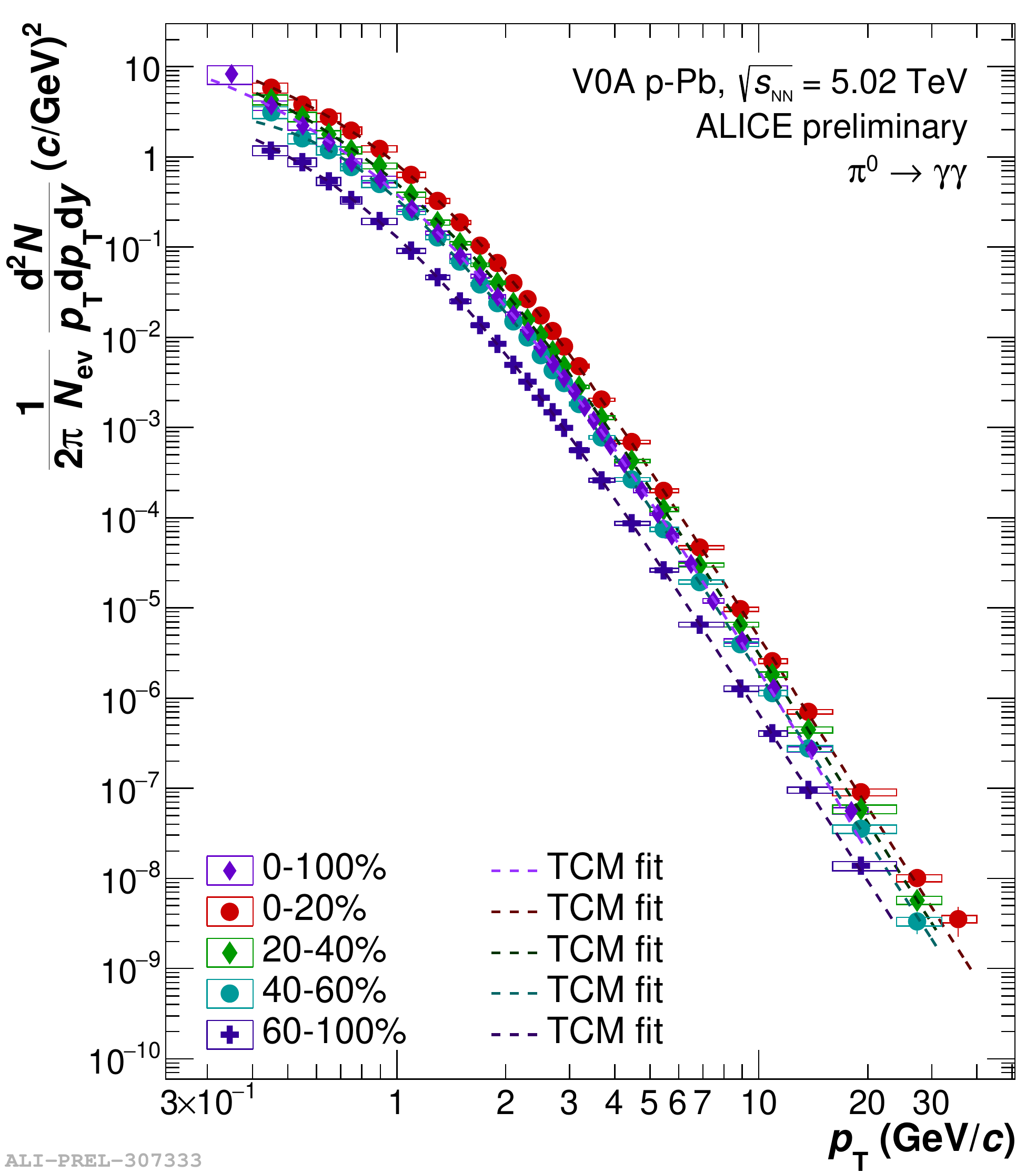}
	\includegraphics[width=0.49\textwidth]{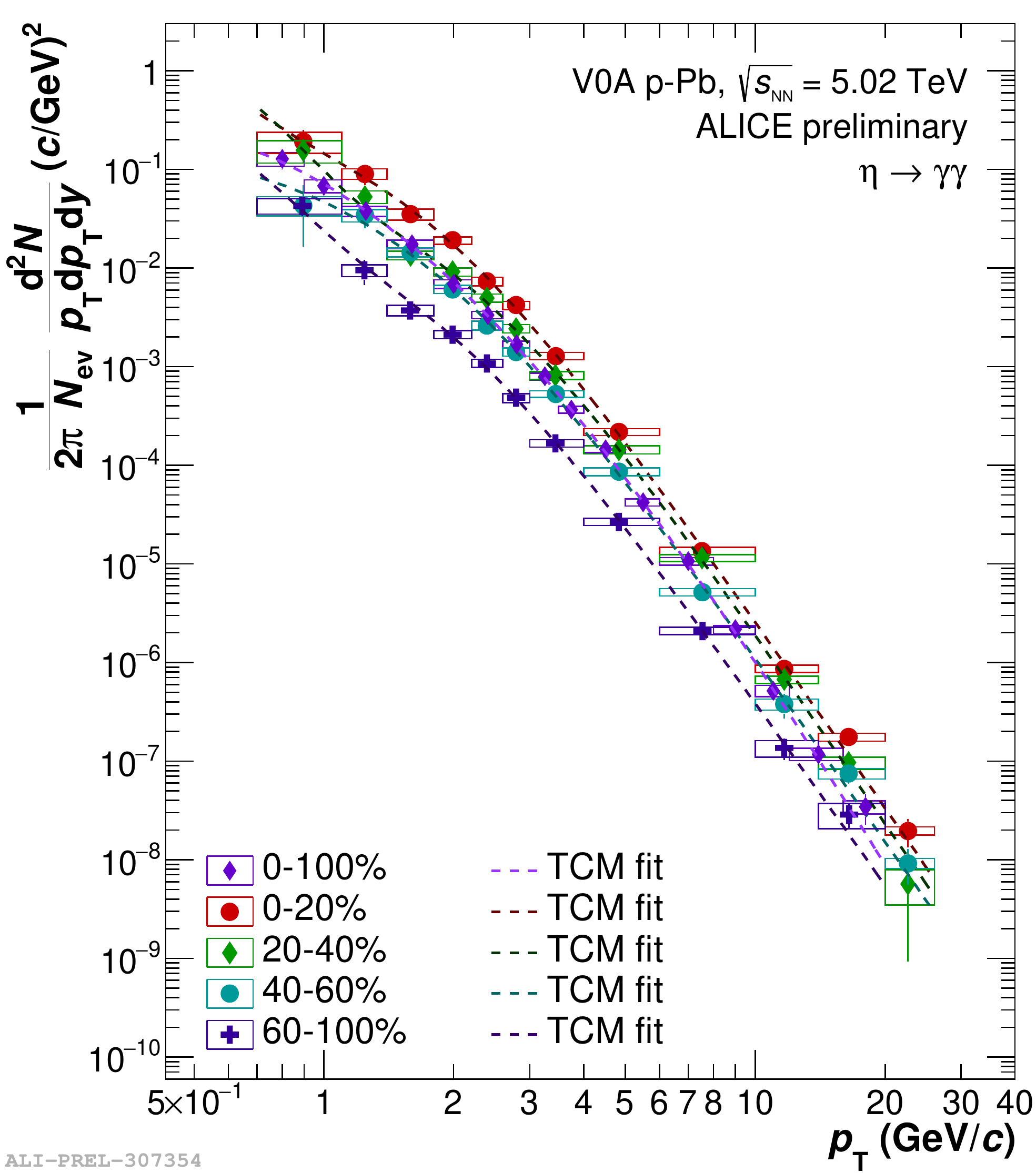}
	\caption{The production of neutral pions (left) and $\eta$ mesons (right), for different collision centralities, in \pPb collisions at \sNN$ = 5.02$ TeV.}
	\label{pPb_pi0_eta_production}
\end{figure}

\begin{figure}
	\centering
	\includegraphics[width=0.49\textwidth]{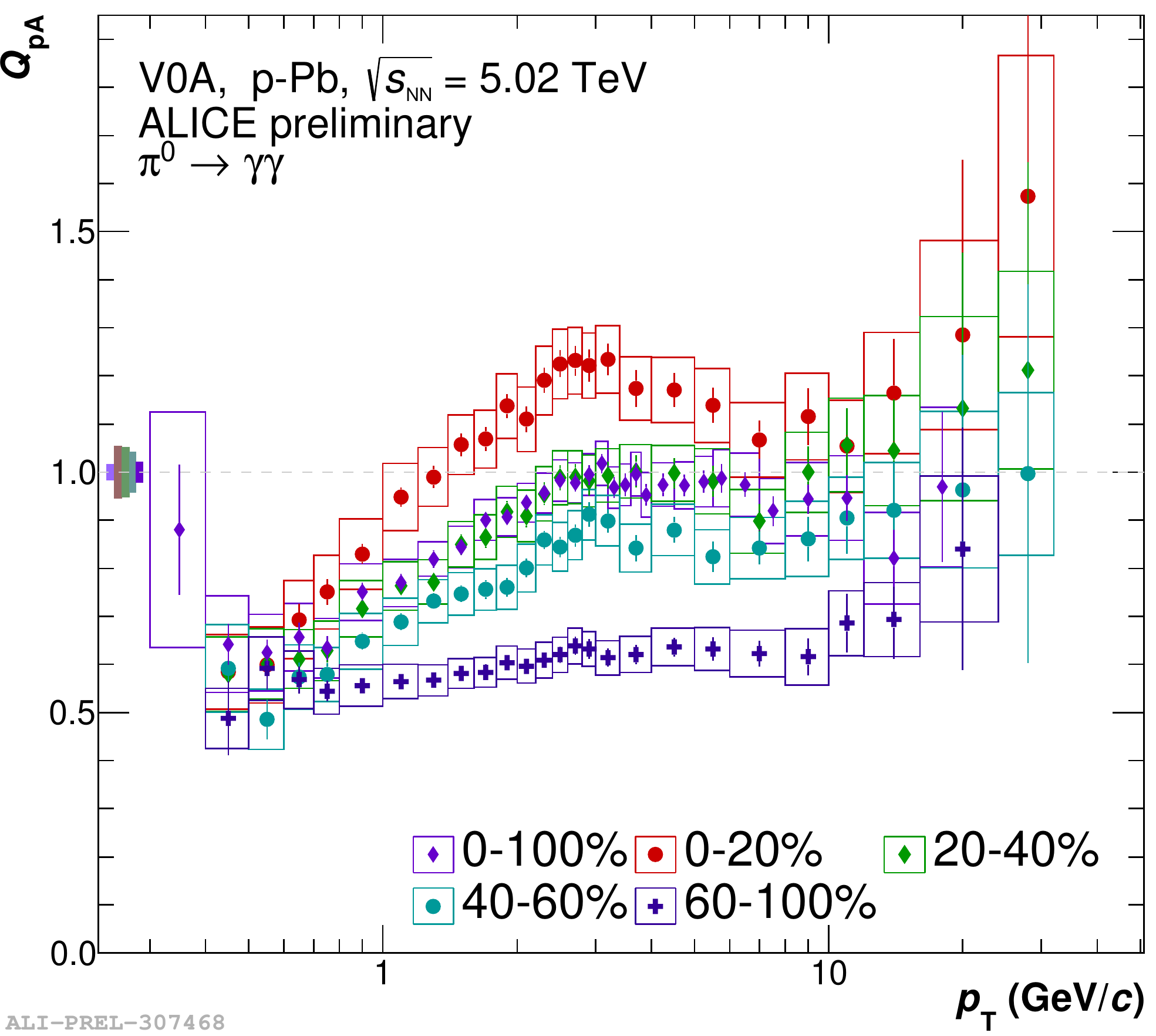}
	\includegraphics[width=0.49\textwidth]{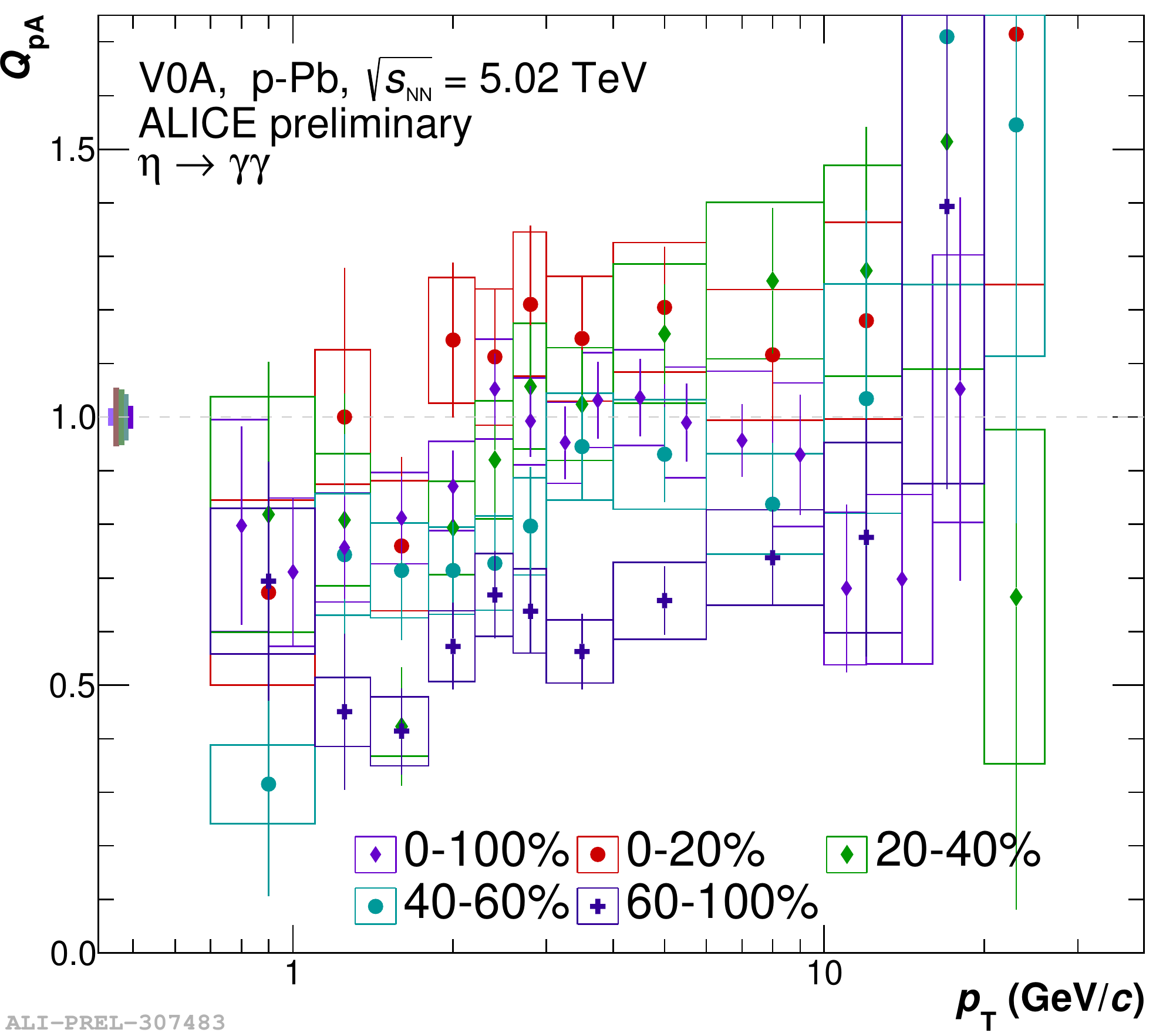}
	\caption{The $Q_{\mathrm{pA}}$ of neutral pions (left) and $\eta$ mesons (right) in \pPb collisions at \sNN$ = 5.02$ TeV.}
	\label{pPb_QpA_pi0_eta}
\end{figure}

In \pPb collisions the spectra of $\pi^{0}$ and $\eta$ mesons are measured for minimum-bias (MB) collisions using the PCM and calorimeters PHOS and EMCal~\cite{AlicepPb:2018}. The invariant yield of neutral pions as function of transverse momentum in \pPb collisions at \sNN$ = 5.02$ TeV is shown in Fig.~\ref{pp_pPb_production} (right).
It covers $0.3<\pT<20$ GeV/$c$, and is compared to various event simulators~\cite{Roesler:2000he} and theoretical predictions, which are mostly consistent with the measurement due to the relatively large uncertainties. Figure \ref{pPb_pi0_eta_production} shows the invariant yield of $\pi^{0}$(left) and $\eta$ meson(right) for different collision centralities, where the V0A detector is used as centrality estimation. A clear ordering is observed; more central \pPb collisions produce more neutral mesons. To study the cold nuclear matter effects the $Q_{\mathrm{pA}} = \frac{dN^{\mathrm{pA}}/d p_{\mathrm{T}}}{<T_{\mathrm{pA}}>d\sigma^{\mathrm{pp}}/d p_{\mathrm{T}}}$ is calculated. The result is shown in Fig. \ref{pPb_QpA_pi0_eta}, with $\pi^{0}$ (left) and $\eta$ meson (right). Within uncertainties both the mesons show a similar centrality dependence. Peripheral \pPb collisions shows a rather flat $Q_{\mathrm{pA}}$ as function of $\pT$, while central \pPb shows a clear enhancement at $\pT = 2$ GeV/$c$.

\subsection{Pb--Pb collisions}

\begin{figure}
	\centering
	\includegraphics[width=0.49\textwidth]{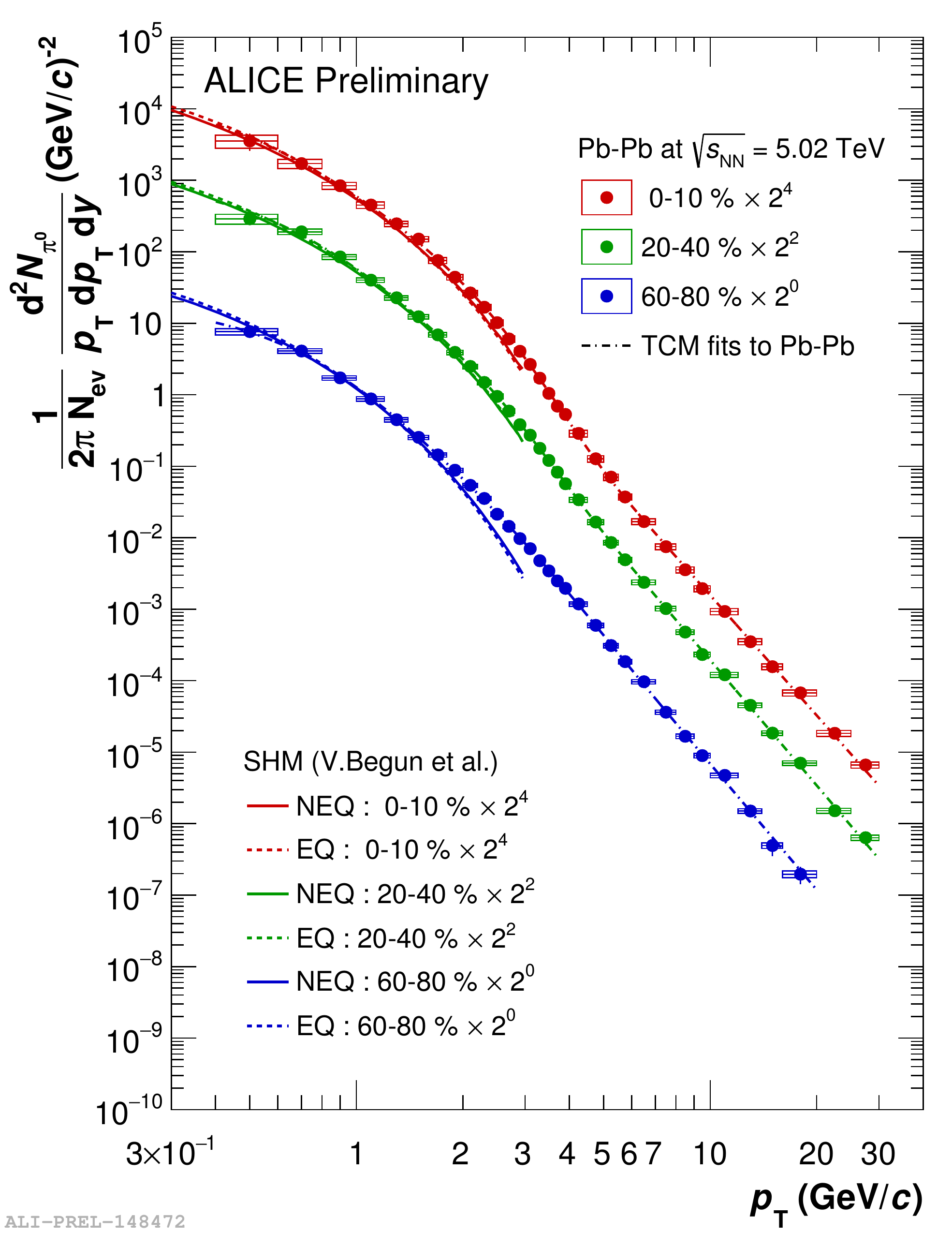}
	\includegraphics[width=0.49\textwidth]{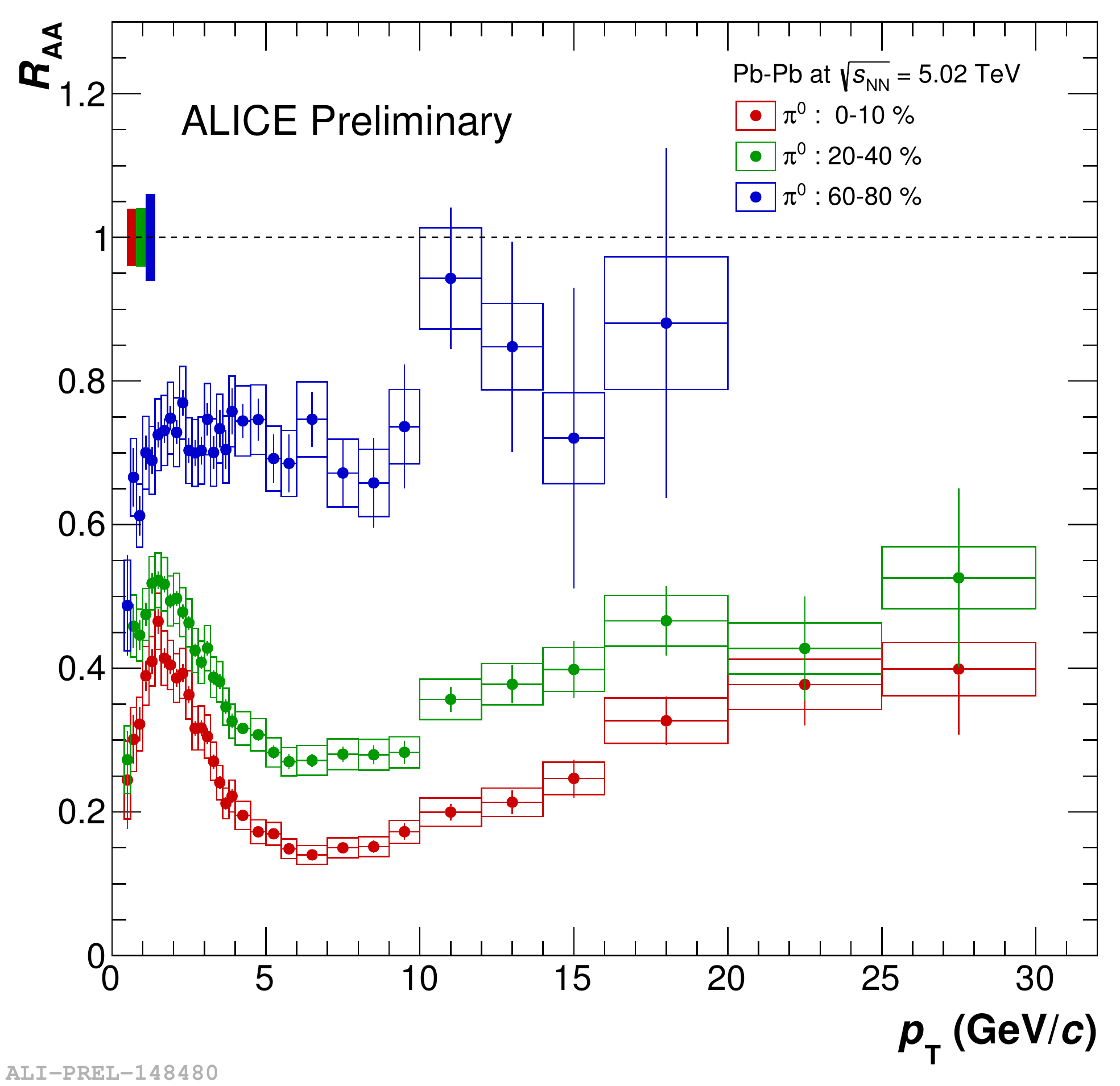}
	\label{PbPb_pi0}
	\caption{The invariant yield of neutral pions (left) and the nuclear modification factor R$_{\mathrm{AA}}$ of neutral pions (right) in \Pb collisions at \sNN$ = 5.02$ TeV.}
\end{figure}

The invariant yield of neutral pions as function of transverse momentum in \Pb collisions at \sNN$ = 5.02$ TeV is shown in Figure \ref{PbPb_pi0} (left), and is measured with the PHOS calorimeter. It covers $0.4<\pT<30$ GeV/$c$ for central and semi-central collisions, and $0.4<\pT<20$ GeV/$c$ for peripheral collisions. Furthermore, the invariant yield is compared to the hydrodynamic SHM model predictions~\cite{Begun:2015}, which describes the yield at lower $\pT$ relatively well, but under-predict the production at higher $\pT$. 
Figure \ref{PbPb_pi0} (right) shows the nuclear modification $R_{\mathrm{AA}} = \frac{dN^{\mathrm{AA}}/d p_{\mathrm{T}}}{<T_{\mathrm{AA}}>d\sigma^{\mathrm{pp}}/d p_{\mathrm{T}}}$ as function of $\pT$. A clear suppression is observed, where central \Pb collisions show more suppression than peripheral ones.

\section{Conclusion}

The neutral meson invariant yield in pp, \pPb, and \Pb collisions has been measured with the ALICE detector. In all these collision systems the measurements are compared to event simulations and model calculations, enabling us to learn about particle production mechanisms. For \pPb and \Pb collisions, the invariant yield is calculated for different collision centralities, thereby probing possible medium effects. In addition, using the neutral meson invariant yield in pp collisions, the nuclear modification factors $Q_{\mathrm{pA}}$ and $R_{\mathrm{AA}}$ are calculated for \pPb and \Pb collisions, respectively.\\
For future measurements, it is planned to utilise all the available photon reconstruction methods and combine all neutral meson measurements into a single precise spectrum, increasing the precision as well as the reach in $\pT$. Furthermore, more differential measurements will help to disentangle the different particle production mechanisms in both pp, \pPb, and \Pb collisions.

\bibliographystyle{JHEP}
\bibliography{biblio.bib}

\end{document}